\documentclass[aps,prb,reprint]{revtex4-1}
\usepackage{amsfonts,amssymb,amsmath,graphicx,color}
\usepackage[colorlinks=true,citecolor=blue,linkcolor=blue,urlcolor=blue]{hyperref}
\newcommand\hc[1]{\hat {\cal #1}}
\newcommand\cc{{\rm c.c.}}
\newcommand\pphantom[1]{\protect\phantom{#1}}
\DeclareMathOperator\tr{tr}
\begin{document}
\title{Theory of spin magnetic quadrupole moment and temperature-gradient-induced magnetization}
\author{Atsuo Shitade}
\affiliation{RIKEN Center for Emergent Matter Science, 2-1 Hirosawa, Wako, Saitama 351-0198, Japan}
\author{Akito Daido}
\affiliation{Department of Physics, Graduate School of Science, Kyoto University, Kyoto 606-8502, Japan}
\author{Youichi Yanase}
\affiliation{Department of Physics, Graduate School of Science, Kyoto University, Kyoto 606-8502, Japan}
\date{\today}
\begin{abstract}
  We revisit a quantum-mechanical formula of the spin magnetic quadrupole moment (MQM) in periodic crystals.
  Two previous attempts were inconsistent with each other; one is gauge dependent, and the other is gauge invariant.
  Here we define the spin MQM by calculating the spin density in a nonuniform system.
  Our definition is analogous to that of the charge polarization, but the result is gauge invariant and coincides with the latter previous one.
  We also formulate what we call gravitomagnetoelectric (gravito-ME) effect, in which the magnetization is induced by a temperature gradient.
  Although the Kubo formula for the gravito-ME effect provides an unphysical divergence at zero temperature,
  we prove that the correct susceptibility is obtained by subtracting the spin MQM from the Kubo formula.
  It vanishes at zero temperature and is related to the ME susceptibility by the Mott relation.
  We explicitly calculate the gravito-ME susceptibility in a Rashba ferromagnet and show its experimental feasibility.
\end{abstract}
\maketitle
\section{Introduction} \label{sec:introduction}
In classical electromagnetism in matter, multipole moments characterize the anisotropy of the charge and magnetization densities,
the latter of which originates from the circulating charge current and spin densities.
Among multipole moments, the magnetic quadrupole moment (MQM) has been believed
to be an important ingredient for the magnetoelectric (ME) effect~\cite{PhysRevB.76.214404,0953-8984-20-43-434203,PhysRevB.88.094429}.
This phenomenon is allowed only when both the inversion and time-reversal symmetries are broken
but enables us to control the charge polarization (CP) by a magnetic field and the magnetization by an electric field.
Note that the magnetization is also induced by an electric field in noncentrosymmetric metals,
which is called inverse spin galvanic or Edelstein effect~\cite{Ivchenko1978,Ivchenko1989,Aronov1989,Edelstein1990233}.
The ME and Edelstein effects are different by their symmetry requirements and mechanisms.
Since the celebrated discovery of the first ME material Cr$_2$O$_3$~\cite{Dzyaloshinskii1960,Astrov1960,Astrov1961,PhysRevLett.6.607,PhysRevLett.7.310},
the MQM and ME effect have been studied in many materials~\cite{Popov1998,JPSJ.74.1419,nature06139,ncomms5796,ncomms3063,ncomms13039}.

When we define multipole moments quantum mechanically in periodic crystals, we suffer from the fact that the position operator is unbounded.
Traditionally, only the atomic multipole moments have been studied in the field of strongly correlated electron systems~\cite{JPSJ.77.064710,RevModPhys.81.807,JPSJ.78.072001}.
Recently, a cluster extension was discussed in the context of the anomalous Hall effect in noncollinear antiferromagnets~\cite{PhysRevB.95.094406}.
However, this difficulty, at least for the dipole moments, was already overcome.
The CP was defined by calculating the charge current density during an adiabatic deformation of the Hamiltonian~\cite{PhysRevB.47.1651,PhysRevB.48.4442,RevModPhys.66.899}.
The result is expressed by the gauge-dependent Berry connection but unique modulo a quantum.
On the other hand, the orbital magnetization (OM) was defined by calculating the grand potential in a magnetic field~\cite{PhysRevLett.99.197202}
and is expressed by the gauge-invariant Berry curvature and magnetic moment.
More recently, the orbital MQM was defined by extending this thermodynamic definition~\cite{PhysRevB.98.020407,PhysRevB.98.060402}.

The spin MQM is classically defined by
\begin{equation}
  M^i_{\pphantom{i} a}
  = \frac{g \mu_{\rm B}}{\hbar} \frac{1}{V} \int {\rm d}^d x x^i s_a({\vec x}), \label{eq:mqm}
\end{equation}
in which $g, \mu_{\rm B}, V, {\vec s}({\vec x})$ are the $g$ factor, Bohr magneton, volume of a $d$-dimensional system, and spin density, respectively.
So far, there were two attempts at a quantum-mechanical theory of the spin MQM in periodic crystals.
One was to calculate the conventional spin current density during an adiabatic deformation~\cite{PhysRevLett.101.077203,PhysRevB.93.195167,1803.01294}.
Similar to the CP, the result is expressed by the Berry connection.
In the absence of spin-orbit interactions (SOIs), this definition seems natural,
because spin is a Noether charge, and the spin MQM is regarded as a spin analog of the CP.
However, in the presence of SOIs, a locally conserved spin current density is no longer well defined, nor is the spin MQM unique even modulo any quanta.
Note that a covariantly conserved spin current density can be defined in some cases~\cite{PhysRevLett.101.106601,PhysRevLett.109.246604,PhysRevLett.118.116801}.
The other attempt was to calculate the grand potential in a nonuniform Zeeman field~\cite{PhysRevB.97.134423}.
The result is gauge invariant and turns out to have a direct relation to the ME susceptibility.
Obviously, these two results are inconsistent with each other; the former is gauge dependent, and the latter is gauge invariant.
One of the goals of this paper is to resolve this inconsistency.

We also shed light on another aspect of multipole moments.
It is well known that the Kubo formulas of the Nernst and thermal Hall conductivities diverge at zero temperature,
and these unphysical results are corrected by adding the OM and twice the heat magnetization (HM),
respectively~\cite{0022-3719-10-12-021,PhysRevB.55.2344,PhysRevLett.106.197202,PhysRevB.84.184406,PhysRevLett.107.236601}.
In fact, Luttinger's gravitational potential~\cite{PhysRev.135.A1505}, introduced to deal with a temperature gradient,
perturbs not only the density matrix but also the charge and heat current densities.
The former yields the Kubo formulas, and the latter turns into the magnetization corrections.
Note that the HM is a heat analog of the OM and characterizes the circulating heat current density.
One of the authors previously introduced a gravitational vector potential by gauging the time translation symmetry
and defined the HM thermodynamicaly with use of a gravitational magnetic field~\cite{Shitade01122014,JPSJ.86.054601}.
More generally, multipole moments may play an important role in such phenomena induced by a temperature gradient.

In particular, the spin magnetization induced by a temperature gradient is of recent interest in the field of spintronics~\cite{WANG20101509,PhysRevB.87.245309,Xiao2016,PhysRevB.98.075307}.
According to the semiclassical Boltzmann theory, the susceptibility vanishes at zero temperature in a two-dimensional electron gas with the Rashba SOI~\cite{WANG20101509,Xiao2016},
where the Edelstein effect and its heat analog are allowed.
On the other hand, in a Rashba ferromagnet, where both the Edelstein and ME effects as well as their heat analogs are allowed,
the Kubo formula of the susceptibility resulted in an unphysical divergence at zero temperature owing to the Fermi-sea terms~\cite{PhysRevB.98.075307}.
This result implies the failure of the Kubo formula for the heat analog of the ME effect and reminds us of the Nernst and thermal Hall effects.
Hence, we expect that a multipole correction is necessary for this phenomenon.

In this paper, we propose another definition of the spin MQM by calculating the spin density in a nonuniform system.
This definition does not rely on an ambiguous spin current density and works in the presence of SOIs.
The obtained formula is gauge invariant and coincides with the thermodynamic formula~\cite{PhysRevB.97.134423}.
If we replace the spin density with the charge density, we reproduce the Berry phase formula of the CP~\cite{PhysRevB.47.1651,PhysRevB.48.4442,RevModPhys.66.899}.
Thus, the spin MQM can be defined both in the same way as the CP and thermodynamically without any inconsistency.
We also discuss the above-mentioned heat analog of the ME effect,
which we call gravito-ME effect because a temperature gradient is described by Luttinger's gravitational electric field~\cite{PhysRev.135.A1505}.
We prove that the spin MQM should be subtracted from the Kubo formula of the gravito-ME susceptibility.
The obtained gravito-ME susceptibility vanishes at zero temperature and is related to the ME susceptibility by the Mott relation.
As a representative, we calculate the gravito-ME susceptibility in a Rashba ferromagnet and show its experimental feasibility.

\section{Thermodynamic formula} \label{sec:thermodynamic}
First, we review the thermodynamic formula of the MQM~\cite{PhysRevB.97.134423,PhysRevB.98.020407,PhysRevB.98.060402}.
We begin with a local thermodynamic relation of the grand potential $\Omega \equiv E - T S - \mu N$,
\begin{equation}
  {\rm d} \Omega
  = -S {\rm d} T - (M_a - \partial_{X^i} M^i_{\pphantom{i} a}) {\rm d} B^a - N {\rm d} \mu, \label{eq:maxwell1}
\end{equation}
in which $S, {\vec M}, N$ are the entropy, spin (orbital) magnetization, and particle number,
and $T \equiv \beta^{-1}, {\vec B}, \mu$ are the temperature, Zeeman (magnetic) field, and chemical potential.
This relation is reasonable when ${\vec B}({\vec X})$ is nonuniform and varies slowly compared with a length scale of the lattice constants.
By integrating by parts, we obtain the thermodynamic definition of the spin (orbital) MQM,
\begin{equation}
  M^i_{\pphantom{i} a}
  \equiv -\frac{\partial \Omega}{\partial (\partial_{X^i} B^a)}. \label{eq:maxwell2}
\end{equation}
This definition yields a direct relation between the spin (orbital) MQM and ME susceptibility for insulators at zero temperature,
\begin{equation}
  -q \frac{\partial M^i_{\pphantom{i} a}}{\partial \mu}
  = \alpha^i_{\pphantom{i} a}, \label{eq:maxwell3}
\end{equation}
with $q$ being the electron charge.

Aiming at practical calculation, we define the auxiliary spin (orbital) MQM by
\begin{equation}
  {\tilde M}^i_{\pphantom{i} a}
  \equiv -\frac{\partial K}{\partial (\partial_{X^i} B^a)}, \label{eq:maxwell4}
\end{equation}
with the energy $K \equiv E - \mu N = \Omega + T S$.
With the help of the Maxwell relations, we can prove a relation between these two quantities,
\begin{equation}
  {\tilde M}^i_{\pphantom{i} a}
  = \frac{\partial (\beta M^i_{\pphantom{i} a})}{\partial \beta}. \label{eq:maxwell5}
\end{equation}
A similar relation is known for the OM~\cite{PhysRevLett.99.197202}.

From now on, we focus on the spin MQM.
To deal with such a slowly varying Zeeman field, we use the gradient expansion of the Keldysh Green's function~\cite{9780521874991}.
See Appendix~\ref{app:gradient} for the details.
Below we consider a clean noninteracting system described by the Hamiltonian ${\cal H}({\vec X}, {\vec p})$.
The first-order perturbation of the energy $K$ with respect to the gradient is expressed by
\begin{align}
  K_D({\vec X})
  = & i \hbar \sum_{n \not= m} \int \frac{{\rm d}^d p}{(2 \pi \hbar)^d}
  \frac{\langle u_n | v^i | u_m \rangle \langle u_m | \partial_{X^i} {\cal H} | u_n \rangle - \cc}{(\epsilon_n - \epsilon_m)^2} \notag \\
  & \times \{f_n (\epsilon_n - \mu) - (\epsilon_n - \epsilon_m) [f_n + f_n^{\prime} (\epsilon_n - \mu)]/2\}. \label{eq:mqm1}
\end{align}
Here we have introduced a complete orthonormal set of wavefunctions $| u_n({\vec X}, {\vec p}) \rangle$ that satisfy
${\cal H}({\vec X}, {\vec p}) | u_n({\vec X}, {\vec p}) \rangle = \epsilon_n({\vec X}, {\vec p}) | u_n({\vec X}, {\vec p}) \rangle$.
$v^i({\vec X}, {\vec p}) \equiv \partial_{p_i} {\cal H}({\vec X}, {\vec p})$ is the velocity operator,
and $f_n({\vec X}, {\vec p}) \equiv f(\epsilon_n({\vec X}, {\vec p}) - \mu)$ with $f(\xi) = (e^{\beta \xi} \mp 1)^{-1}$ being the Bose or Fermi distribution function.

More specifically, we consider ${\cal H}({\vec X}, {\vec p}) = {\cal H}({\vec p}) - (g \mu_{\rm B}/\hbar) {\vec B}({\vec X}) \cdot {\vec s}$.
The first term describes a periodic crystal, and the second term is the perturbation of the nonuniform Zeeman interaction.
$\partial_{X^i} {\cal H}({\vec X}, {\vec p})$ in Eq.~\eqref{eq:mqm1} is equal to $-(g \mu_{\rm B}/\hbar) \partial_{X^i} B^a({\vec X}) \cdot s_a$.
Then, we successfully reproduce the thermodynamic formula of the spin MQM~\cite{PhysRevB.97.134423},
\begin{subequations} \begin{align}
  M^i_{\pphantom{i} a}
  = & \frac{g \mu_{\rm B}}{\hbar} \sum_n \int \frac{{\rm d}^d p}{(2 \pi \hbar)^d} \notag \\
  & \times \left[-\Omega^i_{\pphantom{i} an} \int_{\epsilon_n - \mu}^{\infty} {\rm d} z f(z) + m^i_{\pphantom{i} an} f_n\right], \label{eq:mqm2a} \\
  \Omega^i_{\pphantom{i} an}
  \equiv & i \hbar \sum_{m (\not= n)} \frac{\langle u_n | v^i | u_m \rangle \langle u_m | s_a | u_n \rangle}{(\epsilon_n - \epsilon_m)^2} + \cc, \label{eq:mqm2b} \\
  m^i_{\pphantom{i} an}
  \equiv & -\frac{i \hbar}{2} \sum_{m (\not= n)} \frac{\langle u_n | v^i | u_m \rangle \langle u_m | s_a | u_n \rangle}{\epsilon_n - \epsilon_m} + \cc \label{eq:mqm2c}
\end{align} \label{eq:mqm2}\end{subequations}
This formula is valid for insulators and metals at zero and nonzero temperature.
See Appendix~\ref{app:thermodynamic} for the details.

\section{Spin density in a nonuniform system} \label{sec:nonuniform}
Here we propose an alternative definition of the spin MQM.
As described in Sec.~\ref{sec:introduction}, the spin MQM is interpreted as a spin analog of the CP.
The CP was defined by calculating the charge current density during an adiabatic deformation~\cite{PhysRevB.47.1651,PhysRevB.48.4442,RevModPhys.66.899}.
However, in general cases with SOIs, the spin MQM cannot be defined similarly, because a spin current density is not well defined.
Our definition is based on a relation for the spin density,
\begin{equation}
  M_a^{\rm tot}
  = M_a - \partial_{X^i} M^i_{\pphantom{i} a}, \label{eq:mqm3}
\end{equation}
which is similar to a relation for the charge density,
\begin{equation}
  \rho^{\rm tot}
  = \rho - \partial_{X^i} P^i. \label{eq:cp1}
\end{equation}
The spin MQM and CP can be defined by calculating the spin and charge densities in a nonuniform system, respectively, in a systematic manner.

Below we focus on a clean noninteracting fermion system.
We consider ${\cal H}({\vec X}, {\vec p}, {\vec B}) = {\cal H}({\vec X}, {\vec p}) - (g \mu_{\rm B}/\hbar) {\vec B} \cdot {\vec s}$,
instead of the Hamiltonian ${\cal H}({\vec X}, {\vec p})$.
This additional Zeeman field is introduced only to make expressions simple and set to zero at the end of the derivation.
For insulators at zero temperature, we obtain the spin density up to the first order with respect to the gradient as
\begin{widetext}
\begin{subequations} \begin{align}
  M_a^{\rm tot}({\vec X})
  = & M_a({\vec X}) - \partial_{X^i} M^i_{\pphantom{i} a}({\vec X}), \label{eq:mqm4a} \\
  M_a({\vec X})
  = & M_{0a}({\vec X}) - \partial_{B^a} \left[\frac{i \hbar}{2} \sum_n^{\rm occ} \int \frac{{\rm d}^d p}{(2 \pi \hbar)^d}
  \langle \partial_{p_i} u_n | (\epsilon_n + {\cal H} - 2 \mu) | \partial_{X^i} u_n \rangle + \cc\right], \label{eq:mqm4b} \\
  M^i_{\pphantom{i} a}({\vec X})
  = & -\frac{i \hbar}{2} \sum_n^{\rm occ} \int \frac{{\rm d}^d p}{(2 \pi \hbar)^d}
  \langle \partial_{p_i} u_n | (\epsilon_n + {\cal H} - 2 \mu) | \partial_{B^a} u_n \rangle + \cc \label{eq:mqm4c}
\end{align} \label{eq:mqm4}\end{subequations}
Here we have introduced a complete orthonormal set of wavefunctions $| u_n({\vec X}, {\vec p}, {\vec B}) \rangle$ that satisfy
${\cal H}({\vec X}, {\vec p}, {\vec B}) | u_n({\vec X}, {\vec p}, {\vec B}) \rangle = \epsilon_n({\vec X}, {\vec p}, {\vec B}) | u_n({\vec X}, {\vec p}, {\vec B}) \rangle$.
$M_{0a}({\vec X})$ is the unperturbed spin density, and ${\rm occ}$ represents the summation over the occupied bands.
See Appendix~\ref{app:nonuniform} for the details.

$\mu$ in Eq.~\eqref{eq:mqm4} originates from Eq.~\eqref{eq:nonuniform7} and is an integral constant at first.
In fact, we find
\begin{align}
  \frac{\partial M_a^{\rm tot}({\vec X})}{\partial \mu}
  = & \sum_n^{\rm occ} \int \frac{{\rm d}^d p}{(2 \pi \hbar)^d}
  [\partial_{B^a} (i \hbar \langle \partial_{p_i} u_n | \partial_{X^i} u_n \rangle + \cc) - \partial_{X^i} (i \hbar \langle \partial_{p_i} u_n | \partial_{B^a} u_n \rangle + \cc)] \notag \\
  = & \sum_n^{\rm occ} \int \frac{{\rm d}^d p}{(2 \pi \hbar)^d}
  [\partial_{B^a} (i \hbar \langle \partial_{p_i} u_n | \partial_{X^i} u_n \rangle + \cc) + \partial_{X^i} (i \hbar \langle \partial_{B^a} u_n | \partial_{p_i} u_n \rangle + \cc)
  + \hbar \partial_{p_i} (i \langle \partial_{X^i} u_n | \partial_{B^a} u_n \rangle + \cc)]
  = 0, \label{eq:mqm5}
\end{align}
\end{widetext}
and hence the spin density $M_a^{\rm tot}({\vec X})$ is independent of $\mu$, which is physically reasonable.
Here we have added the total derivative with respect to $p_i$ to the integrand and used the Bianchi identity.
Nonetheless, by identifying $\mu$ as the chemical potential,
we can interpret $M_a({\vec X}) = M_{0a}({\vec X}) - \partial_{B^a} K_D({\vec X})$ as the dipole contribution to the spin density,
in which $K_D({\vec X})$ is the variation of the energy in Eq.~\eqref{eq:mqm1}.
Also,
\begin{equation}
  M^i_{\pphantom{i} a}
  = \frac{g \mu_{\rm B}}{\hbar} \sum_n^{\rm occ} \int \frac{{\rm d}^d p}{(2 \pi \hbar)^d} [\Omega^i_{\pphantom{i} an} (\epsilon_n - \mu) + m^i_{\pphantom{i} an}], \label{eq:mqm6}
\end{equation}
is identical to the thermodynamic formula of the spin MQM in Eq.~\eqref{eq:mqm2a} for insulators at zero temperature.

We readily discuss how the chemical potential dependence of the spin MQM emerges,
which yields the ME effect as indicated by Eq.~\eqref{eq:maxwell3}.
For insulators at zero temperature, the ME susceptibility should not depend on the chemical potential, and hence the spin MQM linearly depends on it.
From Eq.~\eqref{eq:mqm5}, we find that the chemical potential dependence of the spin MQM comes from the dipole moment.
This dipole moment, given in Eq.~\eqref{eq:mqm4b}, may be nonzero only near the surface because of $|\partial_{X^i} u_n \rangle$.

We can also reproduce the well-known formula of the CP~\cite{PhysRevB.47.1651,PhysRevB.48.4442,RevModPhys.66.899}
by replacing $(g \mu_{\rm B}/\hbar) s_a$ with $q$.
For insulators at zero temperature, we obtain the charge density up to the first order with respect to the gradient as
\begin{subequations} \begin{align}
  \rho^{\rm tot}({\vec X})
  = & \rho_0({\vec X}) - \partial_{X^i} P^i({\vec X}), \label{eq:cp2a} \\
  P^i({\vec X})
  = & i \hbar q \sum_n^{\rm occ} \int \frac{{\rm d}^d p}{(2 \pi \hbar)^d} \langle u_n | \partial_{p_i} u_n \rangle. \label{eq:cp2b}
\end{align} \label{eq:cp2}\end{subequations}
See Appendix~\ref{app:cp} for the details.
Thus, the CP defined by calculating the charge density in a nonuniform system agrees with
that by calculating the charge current density during an adiabatic deformation~\cite{PhysRevB.47.1651,PhysRevB.48.4442,RevModPhys.66.899}.

Equation~\eqref{eq:mqm6} is one of our main results.
We have demonstrated that the spin MQM is defined in the same way as the CP.
In contrast to the CP, the formula is gauge invariant and consistent with the thermodynamic formula Eq.~\eqref{eq:mqm2a} in the literature~\cite{PhysRevB.97.134423}.
The former is valid only for insulators at zero temperature,
while the latter is applied to more generic cases, namely, for insulators and metals at zero and nonzero temperature.
Nonetheless, our definition provides a systematic understanding of the CP and spin MQM, the former of which cannot be defined thermodynamically.
The first definition by using an adiabatic deformation~\cite{PhysRevLett.101.077203,PhysRevB.93.195167,1803.01294} has not been reproduced.

\section{Gravito-ME effect} \label{sec:gme}
The ME effect is a phenomenon in which the magnetization is induced by an electric field when both the inversion and time-reversal symmetries are broken.
Here we discuss a heat analog of the ME effect.
The magnetization can also be induced by a temperature gradient, which we call the gravito-ME effect.
The gravito-ME susceptibility $\beta^i_{\pphantom{i} a}$ is defined by $\delta M_a = \beta^i_{\pphantom{i} a} (-\partial_i T)$,
in which $\delta M_a$ is the magnetization measured from its equilibrium value.
In a Rashba ferromagnet, where the above symmetries are broken by the Rashba SOI and Zeeman interaction, respectively,
this susceptibility was calculated by using the Kubo formula~\cite{PhysRevB.98.075307}.
However, the result diverges at zero temperature and hence is unphysical.
Here we prove that the gravito-ME susceptibility is corrected by the spin MQM and related to the ME susceptibility by the Mott relation.

To deal with a temperature gradient, it is useful to introduce Luttinger's gravitational potential that is coupled to the Hamiltonian density~\cite{PhysRev.135.A1505}.
In the presence of the gravitational potential, denoted by $\phi_{\rm g}({\vec X})$, the density matrix is perturbed as $\rho^{\phi_{\rm g}} = \rho + \rho^{\prime}$.
Furthermore, the spin density operator is perturbed as
\begin{equation}
  s_a^{\phi_{\rm g}}({\vec X})
  = [1 + \phi_{\rm g}({\vec X})] s_a({\vec X}). \label{eq:gme1}
\end{equation}
The factor originates from $\sqrt{-g}$ in general relativity, with $g = \det (g_{\mu \nu})$ being the determinant of the metric tensor $g_{\mu \nu}$.
The expectation value of the spin density operator is expressed by
\begin{align}
  \tr [\rho^{\phi_{\rm g}} s_a^{\phi_{\rm g}}({\vec X})]
  = & \tr [\rho s_a({\vec X})] + \tr [\rho^{\prime} s_a({\vec X})] \notag \\
  & + \phi_{\rm g}({\vec X}) \tr [\rho s_a({\vec X})], \label{eq:gme2}
\end{align}
within the first order with respect to the gravitational potential.
The first term is the unperturbed spin density and expressed by
$(g \mu_{\rm B}/\hbar) \tr [\rho s_a({\vec X})] = M_a({\vec X}) - \partial_{X^i} M^i_{\pphantom{i} a}({\vec X})$.
Then, Eq.~\eqref{eq:gme2} is rewritten by
\begin{align}
  (g \mu_{\rm B}/\hbar) \tr [\rho^{\phi_{\rm g}} s_a^{\phi_{\rm g}}({\vec X})]
  = & [1 + \phi_{\rm g}({\vec X})] M_a({\vec X}) \notag \\
  & - \partial_{X^i} \{[1 + \phi_{\rm g}({\vec X})] M^i_{\pphantom{i} a}({\vec X})\} \notag \\
  & + (g \mu_{\rm B}/\hbar) \tr [\rho^{\prime} s_a({\vec X})] \notag \\
  & - [-\partial_{X^i} \phi_{\rm g}({\vec X})] M^i_{\pphantom{i} a}({\vec X}). \label{eq:gme3}
\end{align}
The first and second terms are the dipole and quadrupole contributions to the equilibrium spin density in the presence of the gravitational potential.
As far as the susceptibility is concerned, the magnetization is measured from such equilibrium terms.
The third term yields the Kubo formula.
Since $-\partial_{X^i} \phi_{\rm g}({\vec X})$ in the fourth term corresponds to $-\partial_{X^i} T({\vec X})/T$, the gravito-ME susceptibility is given by
\begin{equation}
  T \beta^i_{\pphantom{i} a}
  = T {\tilde \beta}^i_{\pphantom{i} a} - M^i_{\pphantom{i} a}, \label{eq:gme4}
\end{equation}
in which tilde represents the Kubo formula.
Thus, the gravito-ME susceptibility is corrected by the spin MQM, as the Nernst and thermal Hall conductivities are corrected by the OM and HM,
respectively~\cite{0022-3719-10-12-021,PhysRevB.55.2344,PhysRevLett.106.197202,PhysRevB.84.184406,PhysRevLett.107.236601}.

In the absence of disorder or interactions, the Kubo formulas of the ME and gravito-ME susceptibilities are given by
\begin{subequations} \begin{align}
  \alpha^i_{\pphantom{i} a}
  = & \frac{q g \mu_{\rm B}}{\hbar} \sum_n \int \frac{{\rm d}^d p}{(2 \pi \hbar)^d} \Omega^i_{\pphantom{i} an} f_n, \label{eq:gme5a} \\
  T {\tilde \beta}^i_{\pphantom{i} a}
  = & \frac{g \mu_{\rm B}}{\hbar} \sum_n \int \frac{{\rm d}^d p}{(2 \pi \hbar)^d} \notag \\
  & \times [\Omega^i_{\pphantom{i} an} (\epsilon_n - \mu) + m^i_{\pphantom{i} an}] f_n. \label{eq:gme5b}
\end{align} \label{eq:gme5}\end{subequations}
The spin MQM Eq.~\eqref{eq:mqm2a} and ME susceptibility Eq.~\eqref{eq:gme5a} satisfy the direct relation Eq.~\eqref{eq:maxwell3} for insulators at zero temperature.
The right-hand side of Eq.~\eqref{eq:gme5b} is in general nonzero, and hence ${\tilde \beta}^i_{\pphantom{i} a}$ diverges at zero temeperature.
By subtracting the spin MQM Eq.~\eqref{eq:mqm2a}, we obtain
\begin{align}
  T \beta^i_{\pphantom{i} a}
  = & \frac{g \mu_{\rm B}}{\hbar} \sum_n \int \frac{{\rm d}^d p}{(2 \pi \hbar)^d} \Omega^i_{\pphantom{i} an} \notag \\
  & \times \left[f_n (\epsilon_n - \mu) + \int_{\epsilon_n - \mu}^{\infty} {\rm d} z f(z)\right]. \label{eq:gme6}
\end{align}
This susceptibility vanishes at zero temperature in accordance with physical requirement.
Furthermore, the formulas satisfy the Mott relation at low temperature,
\begin{equation}
  \beta^i_{\pphantom{i} a}
  = \frac{\pi^2 T}{3 q} \frac{\partial \alpha^i_{\pphantom{i} a}}{\partial \mu}(\mu, T = 0), \label{eq:gme7}
\end{equation}
as proved in Appendix~\ref{app:mott}.
The previous study obtained an unphysical result~\cite{PhysRevB.98.075307} because the correction from the spin MQM was overlooked.
In this sense, the spin MQM plays an essential role for the gravito-ME effect.

\begin{figure*}
  \centering
  \includegraphics[clip,width=\textwidth]{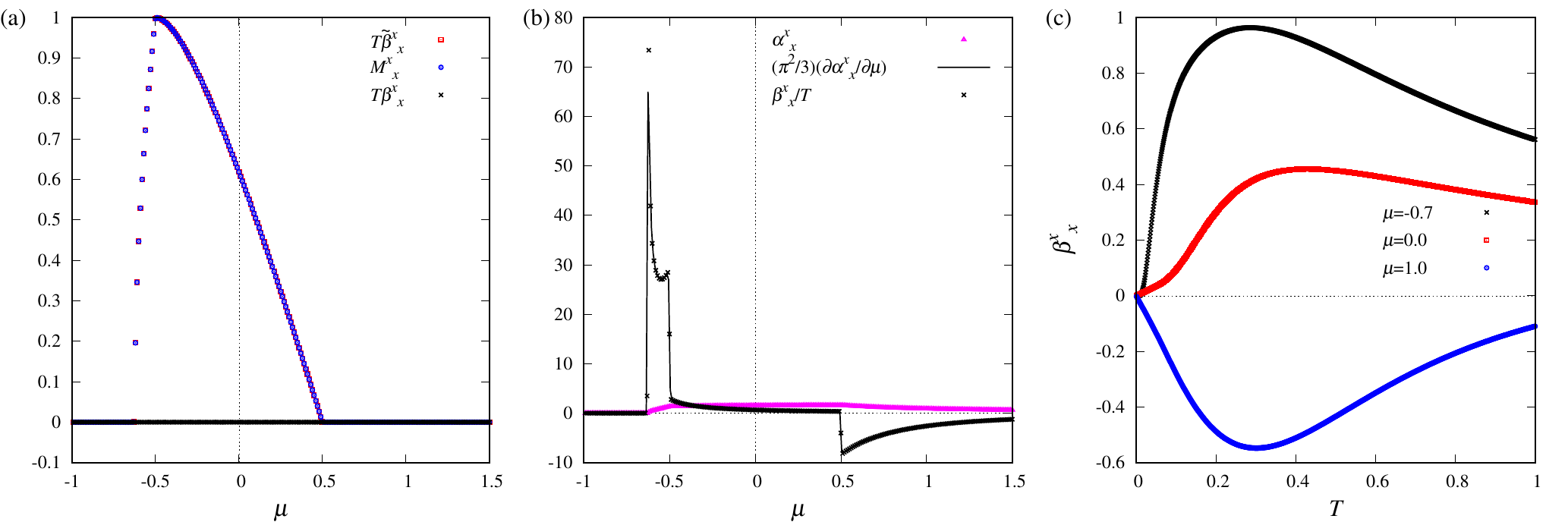}
  \caption{%
  (a) Kubo formula $T {\tilde \beta}^x_{\pphantom{x} x}$, spin MQM $M^x_{\pphantom{x} x}$, and gravito-ME susceptibility $T \beta^x_{\pphantom{x} x}$
  as functions of the chemical potential $\mu$ at $T = 0.001$.
  (b) ME susceptibility $\alpha^x_{\pphantom{x} x}$, its derivative $(\pi^2/3) \partial \alpha^x_{\pphantom{x} x}/\partial \mu$,
  and gravito-ME susceptibility $\beta^x_{\pphantom{x} x}/T$ as functions of the chemical potential $\mu$ at $T = 0.001$.
  (c) Temperature dependence of the gravito-ME susceptibility $\beta^x_{\pphantom{x} x}$ for different values of the chemical potential.
  The units of $\alpha^x_{\pphantom{x} x}, T \beta^x_{\pphantom{x} x}$ are
  $(q/m \alpha^2) (g \mu_{\rm B}/\hbar) (b/8 \pi \alpha), (g \mu_{\rm B}/\hbar) (b/8 \pi \alpha)$, respectively.
  We choose $b = 0.5$ with the energy unit being $m \alpha^2 = 1$.%
  } \label{fig:gme1}
\end{figure*}

\begin{figure*}
  \centering
  \includegraphics[clip,width=\textwidth]{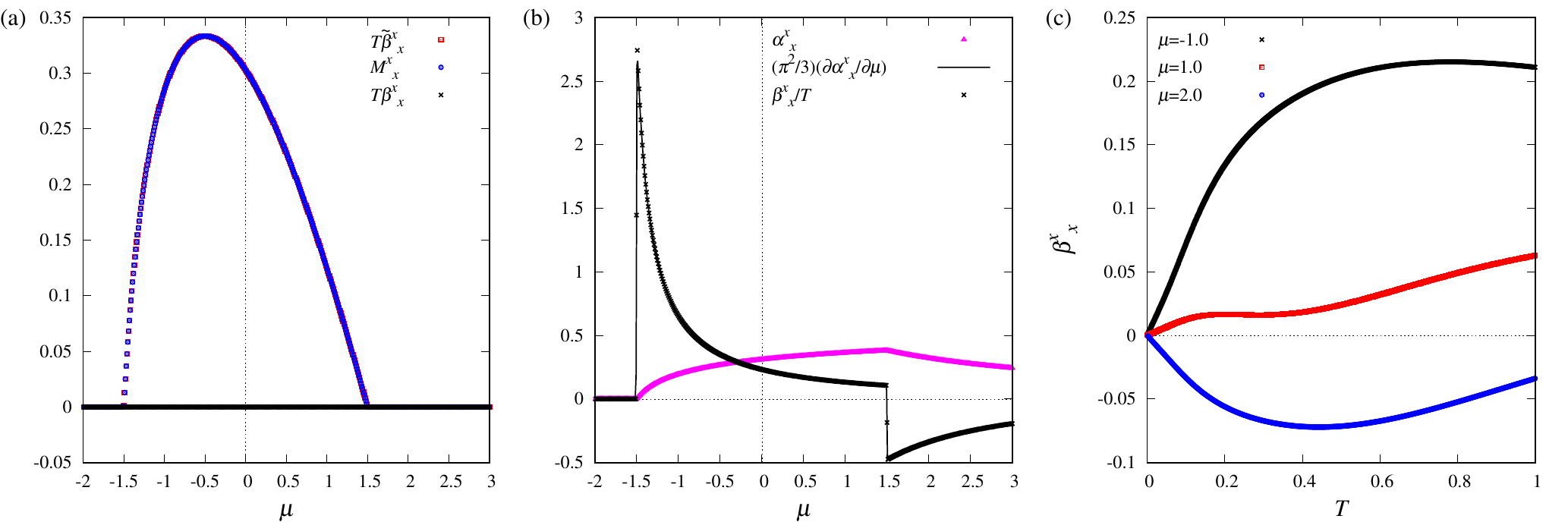}
  \caption{%
  Same as Fig.~\ref{fig:gme1} except for $b = 1.5$.%
  } \label{fig:gme2}
\end{figure*}

As a representative, we consider a Rashba ferromagnet described by
\begin{equation}
  {\cal H}({\vec p})
  = p^2/2 m + \alpha (p_y \sigma_x - p_x \sigma_y) + b \sigma_z, \label{eq:rashba1}
\end{equation}
in which ${\vec \sigma}$ is the Pauli matrix for the spin degree of freedom, i.e., ${\vec s} = (\hbar/2) {\vec \sigma}$.
The dispersions are given by $\epsilon_{\pm}(p) = p^2/2 m \pm \sqrt{(\alpha p)^2 + b^2} \equiv \epsilon(p) \pm \Delta(p)$,
and Eqs.~\eqref{eq:mqm2b} and \eqref{eq:mqm2c} are expressed by
$\Omega^x_{\pphantom{x} x \pm}(p) = \Omega^y_{\pphantom{y} y \pm}(p) = \mp \hbar^2 \alpha b/4 [\Delta(p)]^3,
m^x_{\pphantom{x} x \pm}(p) = m^y_{\pphantom{y} y \pm}(p) = \hbar^2 \alpha b/4 [\Delta(p)]^2$, respectively.
Figure~\ref{fig:gme1} shows the chemical potential and temperature dependences of the ME and gravito-ME susceptibilities for $b = 0.5$.
We set the energy unit to $m \alpha^2 = 1$.
In Fig.~\ref{fig:gme1}(a), the Kubo formula $T {\tilde \beta}^x_{\pphantom{x} x}$ is nonzero even at $T = 0.001$, as already reported~\cite{PhysRevB.98.075307}.
However, the spin MQM $M^x_{\pphantom{x} x}$ is almost equal to $T {\tilde \beta}^x_{\pphantom{x} x}$,
leading to the almost vanishing gravito-ME susceptibility $T \beta^x_{\pphantom{x} x}$.
Figures~\ref{fig:gme1}(b) and (c) clearly show that $\beta^x_{\pphantom{x} x}$ vanishes at zero temperature and satisfies the Mott relation.
In general, the gravito-ME susceptibility is dramatically enhanced at band edges and anticrossing points.
We also show the results for $b = 1.5$ in Fig.~\ref{fig:gme2}.

We believe that the gravito-ME effect is experimentally observable.
In a polar seminconductor BiTeI, the giant Rashba SOI $\hbar \alpha = 3.85~{\rm eV} {\rm \AA}$ was found with use of angle-resolved photoemission spectroscopy~\cite{Ishizaka2011}.
The effective mass is $m = 0.1 m_{\rm e}$, leading to $m \alpha^2 = 2.3 \times 10^3~{\rm K}$.
Although this material is not ferromagnetic, we can apply a magnetic field.
The orbital effect is suppressed for $\omega_{\rm c} \tau \ll 1$, in which $\omega_{\rm c}, \tau$ are the cyclotron frequency and relaxation time, respectively.
Even in such a strongly disordered system, a heat analog of the Edelstein effect is allowed~\cite{WANG20101509,PhysRevB.87.245309,Xiao2016,PhysRevB.98.075307},
but the direction of the induced magnetization is different from that of the gravito-ME effect.
By using the $g$ factor $g = 63$ estimated by quantum oscillations~\cite{1306.1747}, we obtain $b = g \mu_{\rm B} B/2 = 2.1 \times 10^2~{\rm K}$ for $B = 10~{\rm T}$.
$b/m \alpha^2 = 9.1 \times 10^{-2}$ is less than unity.
For $\mu/m \alpha^2 = -0.2$, where the Mott relation holds up to $T/m \alpha^2 = 0.01$,
we find the induced magnetization $M_x/\mu_{\rm B} = 2 \times 10^3 T (\Delta T/L_x) (L_x L_y)$ by the gravito-ME effect.
$\Delta T$ is the temperature difference in the $x$ direction, and $L_x, L_y$ are the sample widths.
We estimate $M_x/\mu_{\rm B} = 8 \times 10^{-2}$ for $T = 20~{\rm K}, \Delta T = 2~{\rm K}, L_y = 1~{\rm \mu m}$.
This value of the magnetization is observable by various experimental methods.
Note that the Rashba ferromagnet is also realized in bilayer devices composed of ferromagnetic and nonmagnetic metals such as Co/Pt~\cite{Miron2010}.
In these systems, the induced magnetization of electrons yields the spin-orbit torque on the magnetization of the ferromagnet~\cite{PhysRevB.77.214429,PhysRevB.78.212405,PhysRevB.79.094422}.

\section{Relation to the spin Hall effect} \label{sec:she}
Finally, we discuss why the first definition of the spin MQM~\cite{PhysRevLett.101.077203,PhysRevB.93.195167,1803.01294} does not work in the presence of SOIs.
To do this, we relate our derivation of the spin MQM to the spin Hall effect.
As is well known, the CP is related to the quantum Hall effect by Thouless's charge pump~\cite{PhysRevB.27.6083,Niu1984}.
The charge current density during an adiabatic deformation is expressed by
\begin{equation}
  J^i(t)
  = -\frac{q}{\hbar} \sum_n \int \frac{{\rm d}^d p}{(2 \pi \hbar)^d} \Omega_{n; p_i t} f_n. \label{eq:qhe1}
\end{equation}
The Berry curvature as well as the magnetic moment, which appears later, in a parameter space $(\lambda_1, \lambda_2, \dots)$ is defined by
\begin{subequations} \begin{align}
  \Omega_{n; \lambda_1 \lambda_2}
  \equiv & i \hbar^2 \langle \partial_{\lambda_1} u_n | \partial_{\lambda_2} u_n \rangle + \cc, \label{eq:qhe2a} \\
  m_{n; \lambda_1 \lambda_2}
  \equiv & -i \hbar^2 \langle \partial_{\lambda_1} u_n | (\epsilon_n - {\cal H}) | \partial_{\lambda_2} u_n \rangle/2 + \cc \label{eq:qhe2b}
\end{align} \label{eq:qhe2}\end{subequations}
If the time dependence in Eq.~\eqref{eq:qhe1} is owing to a vector potential $A_j(t) = -E_j t$,
$\partial_t$ is replaced with $q E_j \partial_{p_j}$, because the Hamiltonian is a function of $p_j - q A_j$.
Thus, we obtain the Hall conductivity,
\begin{equation}
  \sigma^{ij}
  = -\frac{q^2}{\hbar} \sum_n \int \frac{{\rm d}^d p}{(2 \pi \hbar)^d} \Omega_{n; p_i p_j} f_n. \label{eq:qhe3}
\end{equation}
On the other hand, neither a locally conserved spin current density nor a spin pump is well defined in the presence of SOIs.

In our derivation, we have considered a nonuniform system instead of an adiabatic deformation.
As seen in Appendix~\ref{app:cp}, the charge density in a nonuniform system is expressed by
\begin{align}
  \rho^{\rm tot}({\vec X})
  = & \rho_0({\vec X}) + \frac{q}{\hbar} \sum_n \int \frac{{\rm d}^d p}{(2 \pi \hbar)^d} \notag \\
  & \times (\Omega_{n; p_i X^i} f_n + m_{n; p_i X^i} f_n^{\prime}). \label{eq:qhe4}
\end{align}
Provided that the system is nonuniform owing to a vector potential $A_j({\vec X}) = \epsilon_{ijk} B^k X^i/2$,
in which $B^k$ is not a Zeeman field but a magnetic field coupled only to the orbital motion here,
$\partial_{X^i}$ is replaced with $-q \epsilon_{ijk} B^k \partial_{p_j}/2$.
Then, Eq.~\eqref{eq:qhe4} turns into
\begin{align}
  \frac{\partial \rho^{\rm tot}}{\partial B^k}
  = & -\frac{q^2}{\hbar} \frac{1}{2} \epsilon_{ijk} \sum_n \int \frac{{\rm d}^d p}{(2 \pi \hbar)^d} \notag \\
  & \times (\Omega_{n; p_i p_j} f_n + m_{n; p_i p_j} f_n^{\prime}). \label{eq:qhe5}
\end{align}
This is equal to the Fermi-sea term of the Hall conductivity, according to the St\v{r}eda formula~\cite{0022-3719-15-22-005}.
Similarly, the spin density in a nonuniform system, calculated in Appendix~\ref{app:nonuniform}, turns into the spin-orbital susceptibility $\partial M_a^{\rm tot}/\partial B^k$.
If we choose the conserved spin current density ${\rm d} (\{x^i, s_a\}/2)/{\rm d} t$
by adding the torque dipole density $\{x^i, ({\rm d} s_a/{\rm d} t)\}/2$ to the conventional spin current density $\{v^i, s_a\}/2$~\cite{PhysRevLett.96.076604},
the spin-orbital susceptibility is equal to the Fermi-sea term of the spin Hall conductivity~\cite{PhysRevLett.97.236805}.
We believe that the inconsistency regarding the spin MQM can be resolved by calculating the conserved spin current density, but not the conventional one, during an adiabatic deformation.

\section{Summary} \label{sec:summary}
To summarize, we have derived a quantum-mechanical formula of the spin MQM in a periodic crystal in the same way as the CP.
Instead of calculating the conventional spin current density during an adiabatic deformation, which leads to the gauge-dependent result~\cite{PhysRevLett.101.077203},
we have calculated the spin density in a nonuniform system.
Our result Eq.~\eqref{eq:mqm6} is gauge invariant in contrast to the CP and coincides with the thermodynamic formula~\cite{PhysRevB.97.134423}.
The inconsistency in the literature originates from the spin nonconservation in the presence of SOIs,
more precisely, the torque dipole density in the conserved spin current density.
By using the correctly defined spin MQM, we have formulated the temperature-gradient-induced magnetization, namely, the gravito-ME effect.
We have proved that the gravito-ME susceptibility is corrected by the spin MQM as in Eq.~\eqref{eq:gme4}
and related to the ME susceptibility by the Mott relation as in Eq.~\eqref{eq:gme7}.
This phenomenon can be experimentally observed in a Rashba ferromagnet such as BiTeI in a magnetic field and Co/Pt.
Finally, we expect that the susceptibility of any physical quantity to a temperature gradient is corrected by the corresponding multipole moment.
Thus, multipole moments are important not only as microscopic origins of the electromagnetic responses but also in the phenomena induced by a temperature gradient.

\begin{acknowledgments}
  This work was supported by Grants-in-Aid for Scientific Research on Innovative Areas J-Physics (Grant No.~JP15H05884)
  and Topological Materials Science (Grant No.~JP18H04225) from the Japan Society for the Promotion of Science (JSPS),
  and by JSPS KAKENHI (Grants No.~JP15H05745, No.~JP17J10588, No.~JP18H01178, No.~JP18H05227, and No.~JP18K13508).
  A.S. was supported by the RIKEN Special Postdoctoral Researcher Program.
\end{acknowledgments}

\appendix
\section{Gradient expansion} \label{app:gradient}
Gradient expansion is a perturbation theory of the Keldysh Green's function with respect to a slowly varying field~\cite{9780521874991}.
In this method, the Keldysh Green's function ${\hat G}(x_1, x_2)$ is expressed in the Wigner representation,
namely, in terms of the center-of-mass coordinate $X_{12} \equiv (x_1 + x_2)/2$ and relative momentum $p_{12}$.
Since the Wigner transformation is the Fourier transformation with respect to the relative coordinate $x_{12} \equiv x_1 - x_2$,
the Wigner representation of the convolution ${\hat A} \ast {\hat B}(x_1, x_2)$
is expressed by a sort of product of two Wigner representations ${\hat A}(X_{12}, p_{12}) \ast {\hat B}(X_{12}, p_{12})$.
This product, called Moyal product, is noncommutative and approximately given by
\begin{equation}
  {\hat A} \ast {\hat B}
  = {\hat A} {\hat B}
  + (i \hbar/2) (\partial_{X^{\lambda}} {\hat A} \partial_{p_{\lambda}} {\hat B} - \partial_{p_{\lambda}} {\hat A} \partial_{X^{\lambda}} {\hat B}). \label{eq:gradient1}
\end{equation}
Hereafter we use the Wigner represention and drop the arguments $X_{12}, p_{12}$ for simplicity.

The Keldysh Green's function is determined by the Dyson equation,
\begin{equation}
  ({\hc L} - {\hat \Sigma}) \ast {\hat G}
  = {\hat G} \ast ({\hc L} - {\hat \Sigma})
  = 1, \label{eq:gradient2}
\end{equation}
in which ${\hc L}, {\hat \Sigma}$ are the Lagrangian and self-energy.
We expand ${\hat G}, {\hat \Sigma}$ as
\begin{subequations} \begin{align}
  {\hat G}
  = & {\hat G}_0 + (\hbar/2) {\hat G}_D, \label{eq:gradient3a} \\
  {\hat \Sigma}
  = & {\hat \Sigma}_0 + (\hbar/2) {\hat \Sigma}_D. \label{eq:gradient3b}
\end{align} \label{eq:gradient3}\end{subequations}
Here ${\hat G}_0$ is the unperturbed Keldysh Green's function in which effects of disorder or interactions are in principle taken into account.
By substituting Eqs.~\eqref{eq:gradient1} and \eqref{eq:gradient3} into Eq.~\eqref{eq:gradient2}, we obtain ${\hat G}_0 = ({\hc L} - {\hat \Sigma}_0)^{-1}$ and
\begin{equation}
  {\hat G}_D
  = {\hat G}_0 {\hat \Sigma}_D {\hat G}_0
  + i [{\hat G}_0 \partial_{X^{\lambda}} {\hat G}_0^{-1} {\hat G}_0 \partial_{p_{\lambda}} {\hat G}_0^{-1} {\hat G}_0
  - (X^{\lambda} \leftrightarrow p_{\lambda})]. \label{eq:gradient4}
\end{equation}

The Keldysh Green's function contains three independent Green's functions; retarded $G^{\rm R}$, advanced $G^{\rm A}$, and lesser $G^<$.
In particular, $G^<$ is necessary for calculating expectation values.
The unperturbed Green's functions satisfy
\begin{subequations} \begin{align}
  G_0^{\rm R}
  = & ({\cal L} - \Sigma_0^{\rm R})^{-1}, \label{eq:gradient5a} \\
  G_0^{\rm A}
  = & ({\cal L} - \Sigma_0^{\rm A})^{-1}, \label{eq:gradient5b} \\
  G_0^<
  = & \pm (G_0^{\rm R} - G_0^{\rm A}) f(-p_0), \label{eq:gradient5c} \\
  \Sigma_0^<
  = & \pm (\Sigma_0^{\rm R} - \Sigma_0^{\rm A}) f(-p_0), \label{eq:gradient5d}
\end{align} \label{eq:gradient5}\end{subequations}
\begin{widetext}
in which the upper or lower sign represents boson or fermion, and $f(\xi) = (e^{\beta \xi} \mp 1)^{-1}$ is the distribution function at temperature $T = \beta^{-1}$.
By using Eqs.~\eqref{eq:gradient5} and the following ansatz:
\begin{subequations} \begin{align}
  G_D^<
  = & \pm [(G_D^{\rm R} - G_D^{\rm A}) f(-p_0) + G_D^{< (1)} f^{\prime}(-p_0)], \label{eq:gradient6a} \\
  \Sigma_D^<
  = & \pm [(\Sigma_D^{\rm R} - \Sigma_D^{\rm A}) f(-p_0) + \Sigma_D^{< (1)} f^{\prime}(-p_0)], \label{eq:gradient6b}
\end{align} \label{eq:gradient6}\end{subequations}
Eq.~\eqref{eq:gradient4} reads
\begin{subequations} \begin{align}
  G_D^{\rm R}
  = & G_0^{\rm R} \Sigma_D^{\rm R} G_0^{\rm R}
  + i [G_0^{\rm R} \partial_{X^{\lambda}} (G_0^{\rm R})^{-1} G_0^{\rm R} \partial_{p_{\lambda}} (G_0^{\rm R})^{-1} G_0^{\rm R}
  - (X^{\lambda} \leftrightarrow p_{\lambda})], \label{eq:gradient7a} \\
  G_D^{\rm A}
  = & G_0^{\rm A} \Sigma_D^{\rm A} G_0^{\rm A}
  + i [G_0^{\rm A} \partial_{X^{\lambda}} (G_0^{\rm A})^{-1} G_0^{\rm A} \partial_{p_{\lambda}} (G_0^{\rm A})^{-1} G_0^{\rm A}
  - (X^{\lambda} \leftrightarrow p_{\lambda})], \label{eq:gradient7b} \\
  G_D^{< (1)}
  = & G_0^{\rm R} \Sigma_D^{< (1)} G_0^{\rm A}
  - i \{G_0^{\rm R} \partial_{X^0} [(G_0^{\rm R})^{-1} + (G_0^{\rm A})^{-1}] G_0^{\rm A} + \partial_{X^0} (G_0^{\rm R} + G_0^{\rm A})\}. \label{eq:gradient7c}
\end{align} \label{eq:gradient7}\end{subequations}
$G_D^{< (1)}$ is nonzero only when the system is dynamical and depends on $X^0$.
These expressions are valid for disordered or interacting systems as far as a perturbation theory goes.

\section{Derivation of the thermodynamic formula} \label{app:thermodynamic}
Next we derive the thermodynamic formula of the spin MQM with the gradient expansion.
As described in Sec.~\ref{sec:thermodynamic}, we calculate the energy $K \equiv E - \mu N$ instead of the grand potential $\Omega \equiv E - T S - \mu N$.
The energy is expressed by
\begin{equation}
  K(X)
  = \pm i \hbar \int \frac{{\rm d}^D p}{(2 \pi \hbar)^D} \xi \tr (G^<), \label{eq:thermodynamic1}
\end{equation}
in which $D = d + 1$ is the spacetime dimension, and $\xi \equiv -p_0$.
When an external field is static but nonuniform, which is identified as the Zeeman field later,
the first-order perturbation of the energy with respect to the gradient is given by
\begin{align}
  K_D({\vec X})
  = & \pm \frac{i \hbar^2}{2} \int \frac{{\rm d}^D p}{(2 \pi \hbar)^D} \xi \tr (G_D^<) \notag \\
  = & \frac{i \hbar^2}{2} \int \frac{{\rm d}^D p}{(2 \pi \hbar)^D} \xi \tr [(G_D^{\rm R} - G_D^{\rm A}) f(\xi) + G_D^{< (1)} f^{\prime}(\xi)] \notag \\
  = & \frac{i \hbar^2}{2} \int \frac{{\rm d}^D p}{(2 \pi \hbar)^D} f(\xi) \xi \tr (G_D^{\rm R}) + \cc \label{eq:thermodynamic2}
\end{align}
Here we have used Eqs.~\eqref{eq:gradient3a} and \eqref{eq:gradient6a}.
$G_D^{\rm R}$ is readily available from Eq.~\eqref{eq:gradient7a}, and $G_D^{< (1)}$ vanishes.

In the absence of disorder or interactions, Eq.~\eqref{eq:gradient7a} is reduced to
\begin{equation}
  g_D^{\rm R}
  = i g_0^{\rm R} \partial_{X^i} (g_0^{\rm R})^{-1} g_0^{\rm R} \partial_{p_i} (g_0^{\rm R})^{-1} g_0^{\rm R} - (X^i \leftrightarrow p_i), \label{eq:thermodynamic3}
\end{equation}
in which $g_0^{\rm R}({\vec X}, \xi, {\vec p}) = [\xi - {\cal H}({\vec X}, {\vec p}) + \mu + i \eta]^{-1}$ $(\eta \to +0)$
is the retarded Green's function of the Hamiltonian ${\cal H}({\vec X}, {\vec p})$.
By substituting Eq.~\eqref{eq:thermodynamic3} to Eq.~\eqref{eq:thermodynamic2}, we obtain
\begin{equation}
  K_D({\vec X})
  = -\frac{\hbar}{2} \int \frac{{\rm d}^d p}{(2 \pi \hbar)^d} \int \frac{{\rm d} \xi}{2 \pi} f(\xi) \xi
  \tr [g_0^{\rm R} \partial_{X^i} (g_0^{\rm R})^{-1} g_0^{\rm R} \partial_{p_i} (g_0^{\rm R})^{-1} g_0^{\rm R} - (X^i \leftrightarrow p_i)]
  + \cc \label{eq:thermodynamic4}
\end{equation}
Here we introduce a complete orthonormal set of wavefunctions $| u_n({\vec X}, {\vec p}) \rangle$ that satisfy
${\cal H}({\vec X}, {\vec p}) | u_n({\vec X}, {\vec p}) \rangle = \epsilon_n({\vec X}, {\vec p}) | u_n({\vec X}, {\vec p}) \rangle$.
These wavefunctions are replaced with the Bloch wavefunctions later.
By expanding the trace in Eq.~\eqref{eq:thermodynamic4}, we obtain
\begin{align}
  K_D({\vec X})
  = & \frac{\hbar}{2} \sum_{nm} \int \frac{{\rm d}^d p}{(2 \pi \hbar)^d} (\langle u_n | v^i | u_m \rangle \langle u_m | \partial_{X^i} {\cal H} | u_n \rangle - \cc)
  \int \frac{{\rm d} \xi}{2 \pi} f(\xi) \xi [(g_{0n}^{\rm R})^2 g_{0m}^{\rm R} - \cc] \notag \\
  = & -\frac{\hbar}{2} \sum_{n \not= m} \int \frac{{\rm d}^d p}{(2 \pi \hbar)^d} (\langle u_n | v^i | u_m \rangle \langle u_m | \partial_{X^i} {\cal H} | u_n \rangle - \cc) \notag \\
  & \times\int \frac{{\rm d} \xi}{2 \pi} f(\xi) \xi
  \left[\frac{g_{0n}^{\rm R}}{(\epsilon_n - \epsilon_m)^2} - \frac{g_{0m}^{\rm R}}{(\epsilon_m - \epsilon_n)^2} - \frac{(g_{0n}^{\rm R})^2}{\epsilon_n - \epsilon_m} - \cc\right] \label{eq:thermodynamic5}
\end{align}
in which $v^i({\vec X}, {\vec p}) \equiv \partial_{p_i} {\cal H}({\vec X}, {\vec p})$ is the velocity operator,
and $g_{0n}^{\rm R}({\vec X}, \xi, {\vec p}) = (\xi - \epsilon_n({\vec X}, {\vec p}) + \mu + i \eta)^{-1}$.
The intraband process $n = m$ does not contribute to this variation of the energy.
By using $g_{0n}^{\rm R}({\vec X}, \xi, {\vec p}) - \cc = -2 \pi i \delta(\xi - \epsilon_n({\vec X}, {\vec p}) + \mu)$, we can carry out the integral over $\xi$ and obtain
\begin{align}
  K_D({\vec X})
  = & \frac{i \hbar}{2} \sum_{n \not= m} \int \frac{{\rm d}^d p}{(2 \pi \hbar)^d} (\langle u_n | v^i | u_m \rangle \langle u_m | \partial_{X^i} {\cal H} | u_n \rangle - \cc)
  \left[\frac{f_n (\epsilon_n - \mu)}{(\epsilon_n - \epsilon_m)^2} - \frac{f_m (\epsilon_m - \mu)}{(\epsilon_m - \epsilon_n)^2}
  - \frac{f_n + f_n^{\prime} (\epsilon_n - \mu)}{\epsilon_n - \epsilon_m}\right] \notag \\
  = & i \hbar \sum_{n \not= m} \int \frac{{\rm d}^d p}{(2 \pi \hbar)^d}
  \frac{\langle u_n | v^i | u_m \rangle \langle u_m | \partial_{X^i} {\cal H} | u_n \rangle - \cc}{(\epsilon_n - \epsilon_m)^2}
  \{f_n (\epsilon_n - \mu) - (\epsilon_n - \epsilon_m) [f_n + f_n^{\prime} (\epsilon_n - \mu)]/2\}, \label{eq:thermodynamic6}
\end{align}
with $f_n({\vec X}, {\vec p}) \equiv f(\epsilon_n({\vec X}, {\vec p}) - \mu)$.

Finally, we consider a more specific system.
The Hamiltonian ${\cal H}({\vec X}, {\vec p})$ consists of two terms; one is the unperturbed term ${\cal H}({\vec p})$ that describes a periodic crystal,
and the other is the perturbation Zeeman interaction $-(g \mu_{\rm B}/\hbar) {\vec B}({\vec X}) \cdot {\vec s}$.
$\partial_{X^i} {\cal H}({\vec X}, {\vec p})$ in Eq.~\eqref{eq:thermodynamic6} is equal to $-(g \mu_{\rm B}/\hbar) \partial_{X^i} B^a({\vec X}) \cdot s_a$.
Thus, the auxiliary spin MQM Eq.~\eqref{eq:maxwell4} is expressed by
\begin{subequations} \begin{align}
  {\tilde M}^i_{\pphantom{i} a}
  = & \frac{g \mu_{\rm B}}{\hbar} \sum_n \int \frac{{\rm d}^d p}{(2 \pi \hbar)^d}
  \{\Omega^i_{\pphantom{i} an} f_n (\epsilon_n - \mu) + m^i_{\pphantom{i} an} [f_n + f_n^{\prime} (\epsilon_n - \mu)]\}, \label{eq:thermodynamic7a} \\
  \Omega^i_{\pphantom{i} an}
  \equiv & i \hbar \sum_{m (\not= n)} \frac{\langle u_n | v^i | u_m \rangle \langle u_m | s_a | u_n \rangle}{(\epsilon_n - \epsilon_m)^2} + \cc, \label{eq:thermodynamic7b} \\
  m^i_{\pphantom{i} an}
  \equiv & -\frac{i \hbar}{2} \sum_{m (\not= n)} \frac{\langle u_n | v^i | u_m \rangle \langle u_m | s_a | u_n \rangle}{\epsilon_n - \epsilon_m} + \cc \label{eq:thermodynamic7c}
\end{align} \label{eq:thermodynamic7}\end{subequations}
In these expressions, we have replaced $\epsilon_n({\vec X}, {\vec p}), | u_n({\vec X}, {\vec p}) \rangle$
with $\epsilon_n({\vec p}), | u_n({\vec p}) \rangle$ for the unperturbed Hamiltonian ${\cal H}({\vec p})$,
because we have already taken into account the first-order perturbation with respect to $\partial_{X^i} B^a({\vec X})$.
We solve Eq.~\eqref{eq:maxwell5} and reproduce the thermodynamic formula of the spin MQM~\cite{PhysRevB.97.134423} as
\begin{equation}
  M^i_{\pphantom{i} a}
  = \frac{g \mu_{\rm B}}{\hbar} \sum_n \int \frac{{\rm d}^d p}{(2 \pi \hbar)^d}
  \left[-\Omega^i_{\pphantom{i} an} \int_{\epsilon_n - \mu}^{\infty} {\rm d} z f(z) + m^i_{\pphantom{i} an} f_n\right]. \label{eq:thermodynamic8}
\end{equation}
This formula is valid for insulators and metals at zero and nonzero temperature.

\section{Calculation of the spin density in a nonuniform system} \label{app:nonuniform}
We calculate the spin density with the gradient expansion to derive a formula of the spin MQM.
The spin density is expressed by
\begin{equation}
  M_a^{\rm tot}(X)
  = \pm i \hbar \frac{g \mu_{\rm B}}{\hbar} \int \frac{{\rm d}^D p}{(2 \pi \hbar)^D} \tr (s_a G^<). \label{eq:nonuniform1}
\end{equation}
In a nonuniform system, it consists of the unperturbed term $M_{0a}({\vec X})$ and the first-order perturbation with respect to the gradient,
\begin{align}
  M_{Da}({\vec X})
  = & \pm \frac{i \hbar^2}{2} \frac{g \mu_{\rm B}}{\hbar} \int \frac{{\rm d}^D p}{(2 \pi \hbar)^D} \tr (s_a G_D^<) \notag \\
  = & \frac{i \hbar^2}{2} \frac{g \mu_{\rm B}}{\hbar} \int \frac{{\rm d}^D p}{(2 \pi \hbar)^D}
  \tr \{s_a [(G_D^{\rm R} - G_D^{\rm A}) f(\xi) + G_D^{< (1)} f^{\prime}(\xi)]\} \notag \\
  = & \frac{i \hbar^2}{2} \frac{g \mu_{\rm B}}{\hbar} \int \frac{{\rm d}^D p}{(2 \pi \hbar)^D} f(\xi) \tr (s_a G_D^{\rm R}) + \cc \label{eq:nonuniform2}
\end{align}

We evaluate Eq.~\eqref{eq:nonuniform2} in the absence of disorder or interactions.
Instead of the Hamiltonian ${\cal H}({\vec X}, {\vec p})$ that describes the nonuniform system,
we consider ${\cal H}({\vec X}, {\vec p}, {\vec B}) = {\cal H}({\vec X}, {\vec p}) - (g \mu_{\rm B}/\hbar) {\vec B} \cdot {\vec s}$ to make expressions simple later.
This Zeeman field ${\vec B}$ is set to zero at the end of the derivation.
By using Eq.~\eqref{eq:thermodynamic3}, we obtain
\begin{align}
  M_{Da}({\vec X})
  = & -\frac{\hbar}{2} \frac{g \mu_{\rm B}}{\hbar} \int \frac{{\rm d}^d p}{(2 \pi \hbar)^d} \int \frac{{\rm d} \xi}{2 \pi} f(\xi)
  \tr [s_a g_0^{\rm R} \partial_{X^i} (g_0^{\rm R})^{-1} g_0^{\rm R} \partial_{p_i} (g_0^{\rm R})^{-1} g_0^{\rm R} - (X^i \leftrightarrow p_i)] + \cc \notag \\
  = & \frac{\hbar}{2} \frac{g \mu_{\rm B}}{\hbar} \sum_{nmr} \int \frac{{\rm d}^d p}{(2 \pi \hbar)^d}
  (\langle u_n | v^i | u_m \rangle \langle u_m | \partial_{X^i} {\cal H} | u_r \rangle \langle u_r | s_a | u_n \rangle - \cc)
  \int \frac{{\rm d} \xi}{2 \pi} f(\xi) (g_{0n}^{\rm R} g_{0m}^{\rm R} g_{0r}^{\rm R} - \cc) \notag \\
  = & \frac{\hbar}{2} \frac{g \mu_{\rm B}}{\hbar} \sum_{n \not= m \not= r} \int \frac{{\rm d}^d p}{(2 \pi \hbar)^d}
  (\langle u_n | v^i | u_m \rangle \langle u_m | \partial_{X^i} {\cal H} | u_r \rangle \langle u_r | s_a | u_n \rangle - \cc) \notag \\
  & \times \int \frac{{\rm d} \xi}{2 \pi} f(\xi)
  \left[\frac{g_{0n}^{\rm R}}{(\epsilon_n - \epsilon_m) (\epsilon_n - \epsilon_r)} + \frac{g_{0m}^{\rm R}}{(\epsilon_m - \epsilon_n) (\epsilon_m - \epsilon_r)}
  + \frac{g_{0r}^{\rm R}}{(\epsilon_r - \epsilon_n) (\epsilon_r - \epsilon_m)} - \cc\right] \notag \\
  & + \frac{\hbar}{2} \frac{g \mu_{\rm B}}{\hbar} \sum_{n \not= m} \int \frac{{\rm d}^d p}{(2 \pi \hbar)^d}
  (\langle u_n | v^i | u_m \rangle \langle u_m | \partial_{X^i} {\cal H} | u_m \rangle \langle u_m | s_a | u_n \rangle - \cc) \notag \\
  & \times \int \frac{{\rm d} \xi}{2 \pi} f(\xi)
  \left[\frac{g_{0n}^{\rm R}}{(\epsilon_n - \epsilon_m)^2} - \frac{g_{0m}^{\rm R}}{(\epsilon_m - \epsilon_n)^2} + \frac{(g_{0m}^{\rm R})^2}{\epsilon_m - \epsilon_n} - \cc\right] \notag \\
  & - \frac{\hbar}{2} \frac{g \mu_{\rm B}}{\hbar} \sum_{n \not= m} \int \frac{{\rm d}^d p}{(2 \pi \hbar)^d}
  (\langle u_n | v^i | u_m \rangle \langle u_m | \partial_{X^i} {\cal H} | u_n \rangle \langle u_n | s_a | u_n \rangle - \cc) \notag \\
  & \times \int \frac{{\rm d} \xi}{2 \pi} f(\xi)
  \left[\frac{g_{0n}^{\rm R}}{(\epsilon_n - \epsilon_m)^2} - \frac{g_{0m}^{\rm R}}{(\epsilon_m - \epsilon_n)^2} - \frac{(g_{0n}^{\rm R})^2}{\epsilon_n - \epsilon_m} - \cc\right] \notag \\
  & - \frac{\hbar}{2} \frac{g \mu_{\rm B}}{\hbar} \sum_{n \not= r} \int \frac{{\rm d}^d p}{(2 \pi \hbar)^d}
  (\langle u_n | v^i | u_n \rangle \langle u_n | \partial_{X^i} {\cal H} | u_r \rangle \langle u_r | s_a | u_n \rangle - \cc) \notag \\
  & \times \int \frac{{\rm d} \xi}{2 \pi} f(\xi)
  \left[\frac{g_{0n}^{\rm R}}{(\epsilon_n - \epsilon_r)^2} - \frac{g_{0r}^{\rm R}}{(\epsilon_r - \epsilon_n)^2} - \frac{(g_{0n}^{\rm R})^2}{\epsilon_n - \epsilon_r} - \cc\right]. \label{eq:nonuniform3}
\end{align}
These terms represent the interband processes $n \not= m \not= r, n \not= m = r, r = n \not= m, n = m \not= r$.
The intraband process $n = m = r$ does not contribute to the spin density.
We carry out the integrals over $\xi$ and obtain
\begin{align}
  M_{Da}({\vec X})
  = & -\frac{i \hbar}{2} \frac{g \mu_{\rm B}}{\hbar} \sum_{n \not= m \not= r} \int \frac{{\rm d}^d p}{(2 \pi \hbar)^d}
  (\langle u_n | v^i | u_m \rangle \langle u_m | \partial_{X^i} {\cal H} | u_r \rangle \langle u_r | s_a | u_n \rangle - \cc) \notag \\
  & \times \left[\frac{f_n}{(\epsilon_n - \epsilon_m) (\epsilon_n - \epsilon_r)} + \frac{f_m}{(\epsilon_m - \epsilon_n) (\epsilon_m - \epsilon_r)}
  + \frac{f_r}{(\epsilon_r - \epsilon_n) (\epsilon_r - \epsilon_m)}\right] \notag \\
  & - \frac{i \hbar}{2} \frac{g \mu_{\rm B}}{\hbar} \sum_{n \not= m} \int \frac{{\rm d}^d p}{(2 \pi \hbar)^d}
  (\langle u_n | v^i | u_m \rangle \langle u_m | \partial_{X^i} {\cal H} | u_m \rangle \langle u_m | s_a | u_n \rangle - \cc) \notag \\
  & \times \left[\frac{f_n}{(\epsilon_n - \epsilon_m)^2} - \frac{f_m}{(\epsilon_m - \epsilon_n)^2} + \frac{f_m^{\prime}}{\epsilon_m - \epsilon_n}\right] \notag \\
  & + \frac{i \hbar}{2} \frac{g \mu_{\rm B}}{\hbar} \sum_{n \not= m} \int \frac{{\rm d}^d p}{(2 \pi \hbar)^d}
  (\langle u_n | v^i | u_m \rangle \langle u_m | \partial_{X^i} {\cal H} | u_n \rangle \langle u_n | s_a | u_n \rangle - \cc) \notag \\
  & \times \left[\frac{f_n}{(\epsilon_n - \epsilon_m)^2} - \frac{f_m}{(\epsilon_m - \epsilon_n)^2} - \frac{f_n^{\prime}}{\epsilon_n - \epsilon_m}\right] \notag \\
  & + \frac{i \hbar}{2} \frac{g \mu_{\rm B}}{\hbar} \sum_{n \not= r} \int \frac{{\rm d}^d p}{(2 \pi \hbar)^d}
  (\langle u_n | v^i | u_n \rangle \langle u_n | \partial_{X^i} {\cal H} | u_r \rangle \langle u_r | s_a | u_n \rangle - \cc) \notag \\
  & \times \left[\frac{f_n}{(\epsilon_n - \epsilon_r)^2} - \frac{f_r}{(\epsilon_r - \epsilon_n)^2} - \frac{f_n^{\prime}}{\epsilon_n - \epsilon_r}\right] \notag \\
  = & -\frac{i \hbar}{2} \frac{g \mu_{\rm B}}{\hbar} \sum_{n \not= m \not= r} \int \frac{{\rm d}^d p}{(2 \pi \hbar)^d}
  \left[\frac{\langle u_n | v^i | u_m \rangle \langle u_m | \partial_{X^i} {\cal H} | u_r \rangle \langle u_r | s_a | u_n \rangle - \cc}{(\epsilon_n - \epsilon_m) (\epsilon_n - \epsilon_r)}
  + (v^i \rightarrow \partial_{X^i} {\cal H} \rightarrow s_a)\right] f_n \notag \\
  & - \frac{i \hbar}{2} \frac{g \mu_{\rm B}}{\hbar} \sum_{n \not= m} \int \frac{{\rm d}^d p}{(2 \pi \hbar)^d}
  \left[\frac{\langle u_n | v^i | u_m \rangle \langle u_m | \partial_{X^i} {\cal H} | u_m \rangle \langle u_m | s_a | u_n \rangle - \cc}{(\epsilon_n - \epsilon_m)^2}
  + (v^i \rightarrow \partial_{X^i} {\cal H} \rightarrow s_a)\right] f_n \notag \\
  & + \frac{i \hbar}{2} \frac{g \mu_{\rm B}}{\hbar} \sum_{n \not= m} \int \frac{{\rm d}^d p}{(2 \pi \hbar)^d}
  \left[\frac{\langle u_n | v^i | u_m \rangle \langle u_m | \partial_{X^i} {\cal H} | u_n \rangle \langle u_n | s_a | u_n \rangle - \cc}{(\epsilon_n - \epsilon_m)^2}
  + (v^i \rightarrow \partial_{X^i} {\cal H} \rightarrow s_a)\right] \notag \\
  & \times [f_n - (\epsilon_n - \epsilon_m) f_n^{\prime}], \label{eq:nonuniform4}
\end{align}
in which $(v^i \rightarrow \partial_{X^i} {\cal H} \rightarrow s_a)$ represents cyclic permutation.
To make these expressions simple, we use
\begin{subequations} \begin{align}
  \langle u_m | v^i | u_n \rangle
  = & \partial_{p_i} \epsilon_n \delta_{mn} + (\epsilon_n - \epsilon_m) \langle u_m | \partial_{p_i} u_n \rangle, \label{eq:nonuniform5a} \\
  \langle u_m | \partial_{X^i} {\cal H} | u_n \rangle
  = & \partial_{X^i} \epsilon_n \delta_{mn} + (\epsilon_n - \epsilon_m) \langle u_m | \partial_{X^i} u_n \rangle, \label{eq:nonuniform5b} \\
  -(g \mu_{\rm B}/\hbar) \langle u_m | s_a | u_n \rangle
  = & \partial_{B^a} \epsilon_n \delta_{mn} + (\epsilon_n - \epsilon_m) \langle u_m | \partial_{B^a} u_n \rangle. \label{eq:nonuniform5c}
\end{align} \label{eq:nonuniform5}\end{subequations}
Then, by carrying out the summations over $r \not= m (\not= n)$, we obtain
\begin{align}
  M_{Da}({\vec X})
  = & \frac{i \hbar}{2} \sum_n \int \frac{{\rm d}^d p}{(2 \pi \hbar)^d}
  [\langle \partial_{p_i} u_n | Q_n \partial_{X^i} (\epsilon_n + {\cal H}) Q_n | \partial_{B^a} u_n \rangle - \cc + (p_i \rightarrow X^i \rightarrow B^a)] f_n \notag \\
  & - \frac{i \hbar}{2} \sum_n \int \frac{{\rm d}^d p}{(2 \pi \hbar)^d}
  [\langle \partial_{p_i} u_n | (\epsilon_n - {\cal H}) | \partial_{B^a} u_n \rangle \partial_{X^i} \epsilon_n - \cc + (p_i \rightarrow X^i \rightarrow B^a)] f_n^{\prime}. \label{eq:nonuniform6}
\end{align}
$Q_n \equiv 1 - | u_n \rangle \langle u_n |$ is the projection operator, which guarantees the gauge invariance.
Regarding the first term in Eq.~\eqref{eq:nonuniform6}, we find
\begin{align}
  \langle \partial_{p_i} u_n | Q_n \partial_{X^i} (\epsilon_n + {\cal H}) Q_n | \partial_{B^a} u_n \rangle - \cc + (p_i \rightarrow X^i \rightarrow B^a)
  = & \partial_{X^i} [\langle \partial_{p_i} u_n | (\epsilon_n + {\cal H} - 2 \mu) | \partial_{B^a} u_n \rangle - \cc] \notag \\
  & + (p_i \rightarrow X^i \rightarrow B^a), \label{eq:nonuniform7}
\end{align}
in which $\mu$ is not the chemical potential but an integral constant.
For insulators at zero temperature, we can drop the second term in Eq.~\eqref{eq:nonuniform6} and obtain
\begin{subequations} \begin{align}
  M_a^{\rm tot}({\vec X})
  = & M_a({\vec X}) - \partial_{X^i} M^i_{\pphantom{i} a}({\vec X}), \label{eq:nonuniform8a} \\
  M_a({\vec X})
  = & M_{0a}({\vec X}) - \partial_{B^a} \left[\frac{i \hbar}{2} \sum_n^{\rm occ} \int \frac{{\rm d}^d p}{(2 \pi \hbar)^d}
  \langle \partial_{p_i} u_n | (\epsilon_n + {\cal H} - 2 \mu) | \partial_{X^i} u_n \rangle + \cc\right], \label{eq:nonuniform8b} \\
  M^i_{\pphantom{i} a}({\vec X})
  = & -\frac{i \hbar}{2} \sum_n^{\rm occ} \int \frac{{\rm d}^d p}{(2 \pi \hbar)^d}
  \langle \partial_{p_i} u_n | (\epsilon_n + {\cal H} - 2 \mu) | \partial_{B^a} u_n \rangle + \cc \label{eq:nonuniform8c}
\end{align} \label{eq:nonuniform8}\end{subequations}
Here we have added the unperturbed term $M_{0a}({\vec X})$, and ${\rm occ}$ represents the summation over the occupied bands.
If we identify $\mu$ as the chemical potential, we use Eq.~\eqref{eq:nonuniform5} again
and find that $M_a({\vec X}) = M_{0a}({\vec X}) - \partial_{B^a} K_D({\vec X})$ is the dipole contribution to the spin density,
in which $K_D({\vec X})$ is the variation of the energy in Eq.~\eqref{eq:thermodynamic6}.
We also find that
\begin{equation}
  M^i_{\pphantom{i} a}
  = \frac{g \mu_{\rm B}}{\hbar} \sum_n^{\rm occ} \int \frac{{\rm d}^d p}{(2 \pi \hbar)^d} [\Omega^i_{\pphantom{i} an} (\epsilon_n - \mu) + m^i_{\pphantom{i} an}]. \label{eq:nonuniform9}
\end{equation}
is identical to the thermodynamic formula of the spin MQM Eq.~\eqref{eq:thermodynamic8} for insulators at zero temperature.
In this expression, we have replaced $\epsilon_n({\vec X}, {\vec p}, {\vec B}), | u_n({\vec X}, {\vec p}, {\vec B}) \rangle$ with $\epsilon_n({\vec p}), | u_n({\vec p}) \rangle$,
because we have already taken into account the first-order perturbation with respect to the gradient.

\section{Rederivation of the CP} \label{app:cp}
We reproduce the well-known formula of the CP~\cite{PhysRevB.47.1651,PhysRevB.48.4442,RevModPhys.66.899}
by replacing $(g \mu_{\rm B}/\hbar) s_a$ with $q$ in Eq.~\eqref{eq:nonuniform4}.
Since the particle number is diagonal, most of the terms in Eq.~\eqref{eq:nonuniform4} vanish.
The charge density is given by
\begin{align}
  \rho_D({\vec X})
  = & i \hbar q \sum_{n \not= m} \int \frac{{\rm d}^d p}{(2 \pi \hbar)^d}
  \frac{\langle u_n | v^i | u_m \rangle \langle u_m | \partial_{X^i} {\cal H} | u_n \rangle - \cc}{(\epsilon_n - \epsilon_m)^2} [f_n - (\epsilon_n - \epsilon_m) f_n^{\prime}/2] \notag \\
  = & i \hbar q \sum_n \int \frac{{\rm d}^d p}{(2 \pi \hbar)^d}
  \{(\langle \partial_{p_i} u_n | \partial_{X^i} u_n \rangle - \cc) f_n - [\langle \partial_{p_i} u_n | (\epsilon_n - {\cal H}) | \partial_{X^i} u_n \rangle - \cc] f_n^{\prime}/2\}, \label{eq:cp3}
\end{align}
in which we have used Eq.~\eqref{eq:nonuniform5} and carried out the summation over $m (\not= n)$.
Regarding the first term in Eq.~\eqref{eq:cp3},
\begin{equation}
  \langle \partial_{p_i} u_n | \partial_{X^i} u_n \rangle - \cc
  = \partial_{p_i} (\langle u_n | \partial_{X^i} u_n \rangle) - \partial_{X^i} (\langle u_n | \partial_{p_i} u_n \rangle) \label{eq:cp4}
\end{equation}
is used.
For insulators at zero temperature, we can drop the second term in Eq.~\eqref{eq:cp3} and obtain
\begin{subequations} \begin{align}
  \rho^{\rm tot}({\vec X})
  = & \rho_0({\vec X}) - \partial_{X^i} P^i({\vec X}), \label{eq:cp5a} \\
  P^i({\vec X})
  = & i \hbar q \sum_n^{\rm occ} \int \frac{{\rm d}^d p}{(2 \pi \hbar)^d} \langle u_n | \partial_{p_i} u_n \rangle. \label{eq:cp5b}
\end{align} \label{eq:cp5}\end{subequations}

\section{Proof of the Mott relation} \label{app:mott}
Here we prove the Mott relation between the ME and gravito-ME susceptibilities in fermion systems.
The chemical-potential derivative of Eq.~\eqref{eq:gme5a} at zero temperature is given by
\begin{equation}
  \frac{\partial \alpha^i_{\pphantom{i} a}}{\partial \mu}(\mu, T = 0)
  = \frac{q g \mu_{\rm B}}{\hbar} \sum_n \int \frac{{\rm d}^d p}{(2 \pi \hbar)^d} \Omega^i_{\pphantom{i} an} \delta(\mu - \epsilon_n). \label{eq:mott1}
\end{equation}
Then, Eq.~\eqref{eq:gme6} is rewritten by
\begin{align}
  T \beta^i_{\pphantom{i} a}
  = & \frac{1}{q} \int {\rm d} \epsilon \frac{\partial \alpha^i_{\pphantom{i} a}}{\partial \epsilon}(\epsilon, T = 0)
  \left[f(\epsilon - \mu) (\epsilon - \mu) + \int_{\epsilon - \mu}^{\infty} {\rm d} z f(z) \right] \notag \\
  = & \frac{1}{q} \int {\rm d} \epsilon \frac{\partial \alpha^i_{\pphantom{i} a}}{\partial \epsilon}(\epsilon, T = 0)
  \int_{\epsilon - \mu}^{\infty} {\rm d} z [-f^{\prime}(z)] z \notag \\
  = & \frac{1}{q} \int {\rm d} \epsilon [\alpha^i_{\pphantom{i} a}(\epsilon, T = 0) - \alpha^i_{\pphantom{i} a}(\epsilon \to -\infty, T = 0)]
  [-f^{\prime}(\epsilon - \mu)] (\epsilon - \mu)
  = \frac{(\pi T)^2}{3 q} \frac{\partial \alpha^i_{\pphantom{i} a}}{\partial \mu}(\mu, T = 0) + \dots \label{eq:mott2}
\end{align}
in which we have used the Sommerfeld expansion.
\end{widetext}
\end{document}